\documentstyle[12pt,epsfig]{article}
\begin{document}

\begin{center} 
{\Large {\bf Gauge Bosons Masses in the context of the Supersymmetric 
$SU(3)_{C}\otimes SU(3)_{L}\otimes U(1)_{N}$ Model}}
\end{center}

\begin{center}
M. C. Rodriguez  \\
{\it Grupo de F{\'{\i}}sica Te\'{o}rica e Matem\'{a}tica F\'{\i}sica \\
Departamento de F\'{\i}sica  \\
Universidade Federal Rural do Rio de Janeiro - UFRRJ \\
BR 465 Km 7, 23890-000 \\
Serop\'edica, RJ, Brazil, \\
email: marcoscrodriguez@ufrrj.br}
\end{center}

\date{\today}

\begin{abstract}
In this article, we present a detailed study of the masses of all 
gauge bosons, as well as explaing recent experimental data regarding 
the $W$-boson mass presented by the CDF collaboration and even possible 
changes that these data can bring to experimental measurements of the 
masses of $Z$-boson mass in the context of the Minimal Supersymmetric 
$SU(3)_{C}\otimes SU(3)_{L}\otimes U(1)_{N}$ Model. We 
also intend to show a phenomenological analysis of possible 
mixtures of gauge bosons in this model. We will show that 
our numerical predictions for the masses of the physical gauge bosons 
are within the current experimental limits.
\end{abstract}

PACS number(s): 12.60. Cn, 12.60. Jv



\section{Introduction}

Recently, Fermilab's CDF collaboration presented its highly 
precise measurements, with an accuracy of the order of 
$\sim 10^{-4}$, for the mass of the $W$-boson mass 
\cite{cdf}
\begin{equation}
\left( M_{W} \right)_{\rm CDF}= \left(
80.4335 \pm 0.0094 \right) \,\ {\mbox GeV,}
\label{cdfresult}
\end{equation}
this value represents an increase greater than $6$ $\sigma$ 
in relation to the value predicted by the Standard Model (SM) 
\cite{sg,Kronfeld:2010bx}
\begin{equation}
\left( M_{W} \right)_{\rm SM}= \left(
80.3505 \pm 0.0077 \right) \,\ {\mbox GeV.}
\label{smvalue}
\end{equation}
We remember, however, that the average value of experimental 
measurements for this mass, prior to this measurement, provides us 
with the following data \cite{average}
\begin{equation}
\left( M_{W} \right)_{\rm EXP}= \left(
80.4133 \pm 0.0080 \right) \,\ {\mbox GeV.}
\label{average}
\end{equation}
It is clear that, if this result is proven by other 
experimental collaborations, we have a strong indication 
of physics beyond SM. To solve this problem, we must have 
at least two very distinct energy scales: one being 
the SM symmetry breaking mass scale given by: 
$v_{\rm SM}=246$ GeV, while we can think of the other scale 
as being the scale of the break of the lepton number and 
which we will denote as $v_{\rm T}$. 

A strong reason for choosing the 
lepton number break is that it allows us to write the following 
Yukawa interactions
\begin{equation}
{\cal L}_{\Delta ,L}=g^{\nu}_{ij}\bar{L}^{c}_{iL}\Delta L_{jL}+hc,
\label{yukawatermogelmini}
\end{equation}
where $\Delta$ is a scalar in the triplet representation of the group 
$SU(2)$
\begin{equation}
\Delta = \left( 
\begin{array}{cc}
\Delta^{0} & \frac{h^{+}}{\sqrt{2}} \\
\frac{h^{+}}{\sqrt{2}} & H^{++}
\end{array} 
\right) \sim \left({\bf 1},{\bf3}, 2 \right), \,\ 
\langle \Delta \rangle = \frac{1}{\sqrt{2}} \left( 
\begin{array}{cc}
v_{\rm T} & 0 \\
0 & 0
\end{array} 
\right)
\label{tripletogelmini}
\end{equation} 
in parenthesis it appears the transformations properties under 
the respective factors 
$(SU(3)_{C},SU(2)_{L},U(1)_{Y})$. When $\Delta^{0}$ obtains a 
non-zero VEV, we generate the following Majorana mass for neutrions
\footnote{It is the so called type-II Seesaw mechanism 
\cite{seesaw2a,seesaw2b,seesaw2c,seesaw2d,seesaw2e}.} 
\cite{Cribier:2019ckv,bilenkyprehist,bilenky-majorana,bilenky-status}
\begin{equation}
M_{\rm M}\bar{\nu}^{c}_{L}\nu_{L}, \,\ M_{\rm M}=g^{\nu} \frac{v_{\rm T}}{\sqrt{2}}.
\end{equation}
The mass generation mechanism of charged leptons remains the same as presented 
in the SM \cite{sg,Kronfeld:2010bx}.

This new scalar provides different contributions to the masses of the 
$W$-boson and $Z$-boson \cite{carlosmajoron,Pisano:1997hk}
\begin{equation}
M^{2}_{W}= \frac{g^{2}}{4}\left( 
v^{2}_{\rm SM}+2 v^{2}_{\rm T} \right), \,\
M^{2}_{Z}= \frac{g^{2}}{4 \cos^{2} \theta_{W}}\left( 
v^{2}_{\rm SM}+4 v^{2}_{\rm T} \right),
\label{masswezgelmini}
\end{equation}
in addition, it provides us with the following very interesting result
\begin{eqnarray}
\rho &\equiv& \left( 
\frac{M^{2}_{W}}{M^{2}_{Z}\cos^{2}\theta_{W}} \right)= \frac{1+2R}{1+4R}\simeq 1-2R \nonumber \\ 
R&=& \frac{v^{2}_{\rm T}}{v^{2}_{\rm SM}},
\label{rhoparametergelmini}
\end{eqnarray} 
the first line of the above equation together with the 
experimentally known value for $\rho$-parameter \cite{pdg}
\begin{equation}
\rho_{\rm EXP} =0.9998\pm 0.0008,
\label{smexpbosonsector}
\end{equation}
implies that we obtaining the following numerical value for the lepton 
number breakdown scale \cite{carlosmajoron} 
\begin{equation}
2R< 0.0008 \,\ \Rightarrow v_{\rm T}\leq 5.5 \,\ {\mbox GeV.}
\label{newmassscalaeinsm}
\end{equation}

It has recently been proposed that this change in the mass 
of the $W$-boson, which we can define as being expressed by 
the following new parameter
\begin{equation}
\delta M^{2}_{W} \equiv \left[ \left( M_{W} \right)_{\rm CDF} \right]^{2}- 
\left[ \left( M_{W} \right) _{\rm SM} \right]^{2} 
=13.3451 \,\ {\mbox GeV}^{2},
\label{expvalmassdiff}
\end{equation}
can be easily explained by a real triplet Higgs  
boson \cite{Blank:1997qa,Chen:2006pb} and the vaccum 
expectation value, VEV, $v_{\rm T}$, of this new scalar must be of the order of $10$ GeV. Another interesting alternative is 
to introduce complex scalars and in this case the VEV must 
satisfy the following bound $v_{\rm T}\approx 3.2$ GeV 
\cite{evans}. There is also, an interesting proposal to 
explain this anomaly using the Minimal $R$-Symmetric 
extension of the Minimal Superymmetric Standard Model, known as MRSSM \cite{Kribs:2007ac,Diessner:2014ksa,Diessner:2019ebm}, and 
in this case we must respect the following inequality 
$|v_{\rm T}|\leq 4$ GeV \cite{Diessner:2014ksa}.

From the theoretical point of view, the SM cannot be a fundamental theory since it has so many 
questions like that: the number of families do not have an answer in its context. One of 
these  possibilities to solve this problem is that, at energies of a few TeVs, the gauge symmetry may be 
\begin{equation} 
SU(3)_{C}\otimes SU(3)_{L} \otimes U(1)_{N},
\end{equation} 
(or it is more known as M331 for shortness) instead of that of 
the SM \cite{pp,pf}.

Although this model coincides at low energies with the SM it explains 
some fundamental questions that are accommodated, but not explained, in the SM. These 
questions are as follows~\cite{pp}
\begin{enumerate}
\item The family number must be a multiple of three in order to cancel 
anomalies;
\item Why $\sin^{2} \theta_{W}<\frac{1}{4}$ is observed;
\item It is the simplest model that includes bileptons of both types: 
scalars and vectors ones;
\item The models have a scalar sector similar to the two Higgs doublets Model, wchich allow to 
predict the quantization of the electric charge~\cite{pr};
\item It solves the strong $CP$-problem~\cite{pal1};
\item The model has several sources of CP 
violation~\cite{Montero:1998nf,Montero:1998yw,Montero:2005yb}.
\end{enumerate} 
They are interesting  possibilities for the physics at the TeV scale, 
some phenomenological analyses was presented by references 
\cite{331susy1,mcr,Rodriguez:2010tn}.

We will begining the review of the M331 by presenting the quark sector, where we place a familly in the triplet representation of 
$SU(3)_{L}$\footnote{In parenthesis it appears the transformations properties under 
the respective factors $(SU(3)_{C},SU(3)_{L},U(1)_{N})$.} \cite{pp}
\begin{eqnarray}
Q_{1L} &=&\left( \begin{array}{c} 
u_{1} \\ 
d_{1} \\
J          
\end{array} \right)_{L} \sim \left({\bf3},{\bf3},\frac{2}{3}\right) \,\ , 
\label{q1l}
\end{eqnarray}
and the respective singlets are given by
\begin{eqnarray}
u_{1R} &\sim& \left({\bf3},{\bf1},\frac{2}{3}\right),\quad
d_{1R} \sim \left({\bf3},{\bf1},-\frac{1}{3}\right),\quad  
J_{R} \sim \left({\bf3},{\bf1},\frac{5}{3}\right). \nonumber \\
\label{q1r}
\end{eqnarray}

The others two quark generations, we put in the 
antitriplet representation 
of $SU(3)_{L}$ \cite{pp}
\begin{equation}
\begin{array}{cc}  
Q_{2L} =\left( \begin{array}{c} 
d_{2} \\ 
u_{2} \\
j_{1}          \end{array} \right)_{L},\quad  
 
Q_{3L} =\left( \begin{array}{c} 
d_{3} \\ 
u_{3} \\
j_{2}          \end{array} \right)_{L} 
\sim \left({\bf3},{\bf\bar{3}},-\frac{1}{3}\right) , 
\end{array}
\label{q23l}
\end{equation} 
and its singlets fields which we can write as follows
\begin{equation}
u_{2R} \,\ , u_{3R} \sim \left({\bf3},{\bf1},\frac{2}{3}\right),\quad
d_{2R} \,\ , d_{3R} \sim \left({\bf3},{\bf1},- \frac{1}{3}\right), \quad
j_{1R} \,\ , j_{2R} \,\ \sim \left({\bf3},{\bf1},- \frac{4}{3} \right) 
\,\ . 
\label{q23r}
\end{equation}
The fact that one family has a different transformation from the other two, could be a possible explanation for why the third family of quarks is 
much more massive compared to the others two as discussed by reference 
\cite{vicente}.

In order to generate masses for those quarks we have to introduce the following 
scalars fields
\begin{equation} 
\eta =\left( \begin{array}{c} 
\eta^{0} \\ 
\eta^{-}_{1} \\
\eta^{+}_{2}          \end{array} \right) \sim 
({\bf1},{\bf3},0),\quad 
\rho =\left( \begin{array}{c} 
\rho^{+} \\ 
\rho^{0} \\
\rho^{++}          \end{array} \right) \sim 
({\bf1},{\bf3},+1),\quad 
\chi = \left( \begin{array}{c} 
\chi^{-} \\ 
\chi^{--} \\
\chi^{0}          \end{array} \right) \sim 
({\bf1},{\bf3},-1),
\label{3t} 
\end{equation}
using those scalars the Yukawa mass term for the quarks are given by \cite{pp}
\begin{eqnarray}
{\cal L}^{{\mbox Y}}&=&
\bar{Q}_{1L}\sum_{i=1}^{3} \left(
G_{1 i}u_{i R}\eta + 
\tilde{G}_{1 i}d_{i R}\rho \right) + 
\lambda_{J} \bar{Q}_{1L}J_{R}\chi \nonumber \\
&+&\sum_{\alpha =2}^{3}\bar{Q}_{\alpha L}\sum_{i=1}^{3} \left(
F_{\alpha i}u_{i R}\rho^{*} + 
\tilde{F}_{\alpha i}d_{i R}\eta^{*} \right)+ 
\sum_{\alpha =2}^{3}\sum_{\beta =1}^{2}\lambda^{\prime}_{\alpha \beta} \bar{Q}_{\alpha L}j_{\beta R}\chi^{*}+hc. \nonumber \\
\label{yukawaquarks}
\end{eqnarray}
The Glashow-Iliopoulos-Maiani 
mechanism, know as GIM-mechanism, for flavor-changing-neutral-currents 
is presented in the following reference \cite{gimm331}.

In this model, we have the leptons transforming as 
\begin{eqnarray}
L_{iL} &=&\left( \begin{array}{c} 
\nu_{i} \\ 
l_{i} \\
l^{c}_{i}          
\end{array} \right)_{L} \sim ({\bf1},{\bf3},0), \,\ i= 1,2,3.
\label{trip}
\end{eqnarray}
with this triplet of leptons defined above as well the scalars defined by 
Eq.(\ref{3t}), we can construct the following Yukawa term for leptons 
\cite{pp,pf}
\begin{eqnarray}
{\cal L}^{{\mbox Y}}_{\eta}&=&- \frac{1}{2}\sum_{i=1}^{3}\sum_{j=1}^{3} G^{\eta}_{ij}\epsilon_{abd}
\overline{(L^{a}_{iL})^{c}}L^{b}_{jL}\eta^{d}+hc,
\label{yukawaeta}
\end{eqnarray}
the Yukawa parameter $ G^{\eta}_{ij}$, which is a matrix $3 \times 3$, is 
anti-symmetric when exchanging family indices $i,j$ and its eigenvalues have 
the following form \cite{pf}
\begin{equation}
\left( \begin{array}{ccc} 
0 & 0 & 0 \\
0 & M & 0 \\
0 & 0 &- M        
\end{array} \right),
\end{equation}
this means that one charged lepton is massles and the other two are degenerate 
in masses, at least at tree level.

The simplest solution to correct this problem is to introduce the following 
anti-sextet \cite{pf}
\begin{equation}
S = \left( \begin{array}{ccc} 
\sigma^{0}_{1}& 
\frac{h^{+}_{2}}{ \sqrt{2}}& \frac{h^{-}_{1}}{ \sqrt{2}} \\ 
\frac{h^{+}_{2}}{ \sqrt{2}}& H^{++}_{1}& \frac{ \sigma^{0}_{2}}{ \sqrt{2}} \\
 \frac{h^{-}_{1}}{ \sqrt{2}}& 
\frac{\sigma^{0}_{2}}{ \sqrt{2}}&  H^{--}_{2}        
\end{array} \right) \sim ({\bf1},{\bf \bar{6}},0),
\label{antisextetdef}
\end{equation}
with this new scalar, we can consider the following Yukawa term
\begin{eqnarray}
{\cal L}^{{\mbox Y}}_{S}&=&- \frac{1}{2}\sum_{i=1}^{3}\sum_{j=1}^{3} G^{S}_{ij}
\overline{(L_{iL})^{c}}L_{jL}S^{ij}+hc,
\label{yukawaS}
\end{eqnarray}
the Yukawa coupling $G^{S}_{ij}$ is anti-symmetric in to change the 
indices $i$ and $j$ \cite{pf}. Therefore the Yukawa coupling for the charged leptons is given by 
${\cal L}^{{\mbox Y}}={\cal L}^{{\mbox Y}}_{\eta}+{\cal L}^{{\mbox Y}}_{S}$, this term will generate the 
following mass matrix for the charged leptons \cite{331susy1}
\begin{equation}
M^{l}_{ij}=G^{\eta}_{ij} \frac{v_{\eta}}{\sqrt{2}}+G^{S}_{ij} \frac{v_{\sigma_{2}}}{\sqrt{2}}.
\end{equation} 
where $v_{\eta}$ is the VEV of the $\langle \eta^{0} \rangle$, the triplet $\eta$, while $v_{\sigma_{2}}$ is the VEV of
the  $\langle\sigma^{0}_{2}\rangle$, the anti-sextet $S$. If $\langle\sigma^{0}_{1}\rangle \neq 0$ neutrinos get 
Majorana mass term 
\begin{equation}
M^{\nu}_{ij}=G^{S}_{ij} \frac{v_{\sigma_{1}}}{\sqrt{2}},
\end{equation}
and in this situation, still appears a Majoron as discussed in 
reference \cite{Montero:1999mc}. However, if $v_{\eta}=0$ then $M^{l} \propto M^{\nu}$ and it would imply that the 
Pontecorvo-Maki-Nakagawa-Sakata (PMNS) matrix obey \cite{bilenkyprehist}
\begin{equation}
V_{PMNS}=(V^{l}_{L})^{\dagger}V^{\nu}_{L}=I.
\end{equation}

We can decompose our anti-sextet scalar field $S$ in 
$SU(3)\otimes SU(2)\otimes U(1)$ representations\footnote{For a better understand our 
notation, we recomend that you review our Eqs.(\ref{tripletogelmini},\ref{antisextetdef}).} in the following way \cite{Montero:1999mc,Liu:1993gy,Liu:1993fwa,DeConto:2015eia}
\begin{eqnarray}
\left( {\bf 1}, {\bf \bar{6}}, 0 \right) \Rightarrow 
\left( {\bf 1}, {\bf 3}, 2 \right) \oplus 
\left( {\bf 1}, {\bf 2}, -1 \right) \oplus 
\left( {\bf 1}, {\bf 1}, -4 \right)
\end{eqnarray}
where
\begin{eqnarray}
S&=& \left( 
\begin{array}{cc} 
T & \frac{\Phi_{S}}{\sqrt{2}} \\ 
\frac{\Phi^{T}_{S}}{\sqrt{2}} & H^{--}_{2}          
\end{array} \right) \sim ({\bf 1}, {\bf \bar{6}}, 0) \nonumber \\
T&=&\left( \begin{array}{cc} 
\sigma^{0}_{1} & \frac{h^{+}_{2}}{\sqrt{2}} \\ 
\frac{h^{+}_{2}}{\sqrt{2}} & H^{++}_{1} \\
\end{array} \right)
\sim ({\bf 1},{\bf 3},2) \nonumber \\
\Phi_{S}&=& \left( 
\begin{array}{c} 
h^{-}_{1} \\ 
\sigma^{0}_{2}          
\end{array} \right)
\sim ({\bf 1},{\bf 2},-1), \quad 
H^{--}_{2}
\sim ({\bf 1},{\bf 1},-4).
\label{sig1:triplet} 
\end{eqnarray}
where $T$ is the Gelmini-Rocandelli $SU(2)$ triplet
\footnote{This triplet was considered in the studies 
presented in the references
\cite{carlosmajoron,Pisano:1997hk} 
to obtain the result presented by 
Eq.(\ref{newmassscalaeinsm}).} \cite{Pisano:1997hk,Gelmini:1980re}, while 
$\Phi_{S}$ is the Higgs doublet of SM and $H^{--}_{2}$ appear in 
the model of Babu presented in reference \cite{babuh2mm}. 

The Gelmini-Rocandelli  mechanism was implemented in M331 
model and the results can be summarized by the following algebric 
expressions \cite{carlosmajoron}
\begin{eqnarray}
\rho&=&\frac{1+r}{1+2r}, \,\ 
\sqrt{r}= \frac{v_{\rm TS}}{\tilde{v}}, \,\ 
\tilde{v}=\sqrt{v^{2}_{\eta}+ v^{2}_{\rho}+ v^{2}_{\rm DS}}, \nonumber \\
M^{2}_{W}&\approx& \frac{g^{2}}{4}\left( 
v^{2}_{\eta}+ v^{2}_{\rho}+ v^{2}_{\rm DS}+2 v^{2}_{\rm TS} 
\right) \,\ \Rightarrow \,\  
\delta M^{2}_{W}= \frac{g^{2}}{2}v^{2}_{\rm TS} \nonumber \\
M^{2}_{Z}&\approx& \frac{g^{2}}{4 \cos^{2} \theta_{W}}\left( 
v^{2}_{\eta}+ v^{2}_{\rho}+ v^{2}_{\rm DS}+4 v^{2}_{\rm TS} 
\right) \,\ \Rightarrow \nonumber \\ 
\delta M^{2}_{Z}&\equiv& \left[ \left( M_{Z} \right)_{\rm CDF} \right]^{2}- 
\left[ \left( M_{Z} \right)_{\rm EXP} \right]^{2} = 
\frac{g^{2}}{4 \cos^{2} \theta_{W}}(4v^{2}_{\rm TS}), 
\nonumber \\
\label{m331anomalia}
\end{eqnarray}
where $v_{\eta}, v_{\rho}$ and $v_{DS}$ denote the VEV of 
doublet of $SU(2)$ and $v_{TS}$ is the VEV of the triplet 
of $SU(2)$. Before we continue, we can rewrite our 
Eq.(\ref{m331anomalia}) as follows
\begin{eqnarray}
\delta M^{2}_{Z}&=&
\frac{2}{\cos^{2} \theta_{W}}(\delta M^{2}_{W})  
\Rightarrow 
\delta M^{2}_{Z}=17.187592 \,\ {\mbox (GeV)}^{2}.
\label{dmzdmwm331}
\end{eqnarray} 
Comparing the Eq.(\ref{expvalmassdiff}) with 
Eq.(\ref{dmzdmwm331}), we obtain the following interesting result
\begin{eqnarray}
\delta M^{2}_{Z} \neq \delta M^{2}_{W},
\label{dmzdmwm331num}
\end{eqnarray} 
on the other hand, the experimental average of the measurements of this gauge boson is \cite{average}
\begin{equation}
\left( M_{Z} \right)_{\rm EXP}= \left(
91.1875 \pm 0.0021 \right) \,\ {\mbox GeV,}
\label{averagemz}
\end{equation}
which implies that the experimental value of the boson mass 
must be changed to the following value
\begin{equation}
\left( M_{Z} \right)_{\rm CDF}=91.19 \,\ {\mbox GeV.}
\label{mzcdfvaluesm331}
\end{equation}

Using the first relation given in 
Eq.(\ref{m331anomalia}) together with 
Eq.(\ref{smexpbosonsector}), we get the following upper 
limit \cite{carlosmajoron}
\begin{equation}
v_{\rm TS}\leq 3.89 \,\ {\mbox GeV}, \,\ 
\tilde{v}^{2}= 
\left( \frac{246}{2} \right)^{2} \,\ {\mbox (GeV)}^{2}.
\label{limitrhom331}
\end{equation}
By another hand, from the second relation given in 
Eq.(\ref{m331anomalia}), we can write
\begin{equation}
v_{\rm TS}= \frac{\sqrt{2} \delta M^{2}_{W}}{g} \simeq 7.9 \,\ {\mbox GeV.}
\label{limitMWm331}
\end{equation}
As conclusion, we can not solve both problem in this model 
under this hypotesys. The solution to the anomaly for the 
mass of the $W$-boson has another hypothesis different 
from this one, and it was recently presented in the 
following reference \cite{VanLoi:2022eir}. 

The minimal supersymmetric $SU(3)_{C}\otimes SU(3)_{L}\otimes U(1)_{N}$ model, 
or it is more known as MSUSY331 for shortness, was considered some years ago 
in the following references \cite{331susy1,mcr,ema1,pal2}. In 
this model beyond 
the anti-sextet $S$, see Eq.(\ref{antisextetdef}), we need to 
introduce a Sextet,  $S^{\prime}$, in the scalar sector to  
cancel chiral anomalies generated by the superpartners of 
$S$ in similar way as we have to add a new doublet scalar 
at Minimal Supersymmetric\footnote{For those interested in the history 
of Supersymmetry see \cite{volkov1,Likhtman:2001st,fay1,marcoshist}, while anyone interested, how to work with supersymmetry see the great review article \cite{ogievetski}.} Standard Model (MSSM)\footnote{The current status of the search for supersymmetry is presented in reference \cite{gladkaza}.} \cite{mcrphysics,kazakov1,Rodriguez:2019mwf}. In this model, as in the SM, neutrinos are massless. If 
we want to give masss to the neutrinos we must break $R$-Parity 
\cite{Montero:2001ch,VicenteMontesinos:2011pf,Rodriguez:2022gfq}. In this 
mechanism, the masses generated are a combination of 
type-I \cite{seesaw1a,seesaw1b,seesaw1c} 
and type-III seesaw \cite{seesaw3} mechanism. 

Therefore, in the MSUSY331 we will have both scalar fields in the triplet 
and anti-triplet representation of the $SU(2)$ group. When their 
neutral component fields, namely the fields 
$\sigma^{0}_{1}$ and $\sigma^{\prime 0}_{1}$ get VEV non-zero we can expain the shift on the 
$W$-mass, defined in Eq.(\ref{expvalmassdiff}), as we will shown in 
this article. We also intend to calculate the masses of the new gauge 
bosons in this model.

We begin this article by presenting the Minimal Supersymmetric 
$SU(3)_{C}\otimes SU(3)_{L}\otimes U(1)_{N}$ Model. 
Our numerical results for the masses of all gauge bosons are presented 
via plots and tables in Sec.(\ref{wcdf}). We finish this article by 
showing our conclusions in the last Section. The analytical expressions 
for the boson mixtures are studied in the first two appendices and 
finally we succinctly show a preliminar numerical analysis of the neutral boson 
mixture.

\section{Minimal Supersymmetric $SU(3)_{C}\otimes SU(3)_{L}\otimes U(1)_{N}$ Model, (MSUSY331).} 
\label{sec:analitico}

We will introduce the following the chiral superfields associated with 
quarks and leptons:
$\hat{L}_{1,2,3}$, $\hat{Q}_{1,2,3}$, $\hat{u}^{c}_{1,2,3}$, 
$\hat{d}^{c}_{1,2,3}$, $\hat{J}^{c}$ and $\hat{j}^{c}_{1,2}$ \cite{331susy1,mcr}. 
The particle content of each chiral superfield and anti-chiral supermultiplet 
is presented in the Tabs.(\ref{lfermionnmssm},\ref{rfermionnmssm}), 
respectivelly. 

\begin{table}[h]
\begin{center}
\begin{tabular}{|c|c|c|}
\hline 
$\mbox{ Chiral Superfield} $ & $\mbox{ Fermion} $ & $\mbox{ Scalar} $ \\
\hline
$\hat{L}_{iL}=( \hat{\nu}_{i}, \hat{l}_{i}, \hat{l}^{c}_{i})^{T}_{L}
\sim({\bf 1},{\bf3},0)$ & 
$L_{iL}=(\nu_{i},l_{i},l^{c}_{i})^{T}_{L}$ & 
$\tilde{L}_{iL}=( \tilde{\nu}_{i}, \tilde{l}_{i}, \tilde{l}^{c}_{i})^{T}_{L}$ \\
\hline
$\hat{Q}_{1L}=(\hat{u}_{1}, \hat{d}_{1}, \hat{J})^{T}_{L}\sim
({\bf 3},{\bf3},(2/3))$ & 
$Q_{1L}=(u_{1},d_{1},J)^{T}_{L}$ & 
$\tilde{Q}_{1L}=(\tilde{u}_{1}, \tilde{d}_{1}, \tilde{J})^{T}_{L}$ \\ 
\hline
\end{tabular}
\end{center}
\caption{\small Particle content in the chiral superfields in MSUSY331 
and we neglected the color indices and see our Eqs.(\ref{q1l},\ref{trip}) 
and $i=1,2,3$.}
\label{lfermionnmssm}
\end{table}

\begin{table}[h]
\begin{center}
\begin{tabular}{|c|c|c|}
\hline 
$\mbox{ Anti-Chiral Superfield} $ & $\mbox{ Fermion} $ & $\mbox{ Scalar} $ \\
\hline
$\hat{Q}_{\alpha L}=(\hat{d}_{\alpha}, \hat{u}_{\alpha}, \hat{j}_{m})^{T}_{L}
\sim({\bf 3},{\bf\bar{3}},-(1/3))$ & 
$Q_{\alpha L}=(d_{\alpha},u_{\alpha},j_{m})^{T}_{L}$ & 
$\tilde{Q}_{\alpha L}=
(\tilde{d}_{\alpha}, \tilde{u}_{\alpha}, \tilde{j}_{m})^{T}_{L}$ \\
\hline
$\hat{u}^{c}_{iL}\sim({\bf \bar{3}},{\bf1},-(2/3))$ & $u^{c}_{iL}\equiv \bar{u}_{iR}$ & 
$\tilde{u}^{c}_{iL}$ \\ 
\hline
$\hat{d}^{c}_{iL}\sim({\bf \bar{3}},{\bf1},(1/3))$ & 
$d^{c}_{iL}\equiv \bar{d}_{iR}$ & 
$\tilde{d}^{c}_{iL}$ \\ 
\hline
$\hat{J}^{c}_{L}\sim({\bf \bar{3}},{\bf1},-(5/3))$ & 
$J^{c}_{L}\equiv \bar{J}_{R}$ & 
$\tilde{J}^{c}_{L}$   \\
\hline
$\hat{j}^{c}_{\beta L}\sim({\bf \bar{3}},{\bf1},(4/3))$ & 
$j^{c}_{\beta L}\equiv \bar{j}_{\beta R}$ & 
$\tilde{j}^{c}_{\beta L}$   \\
\hline
\end{tabular}
\end{center}
\caption{\small Particle content in the anti-chiral superfields in 
MSUSY331 and see our Eqs.(\ref{q1r},\ref{q23l},\ref{q23r}) 
and $\alpha =2,3$, $i=1,2,3$ and $\beta =1,2$.}
\label{rfermionnmssm}
\end{table}

.

The scalars in this model are given by \cite{331susy1,mcr}
\begin{eqnarray}
\eta &=& 
\left( \begin{array}{c} 
\eta^{0} \\ 
\eta^{-}_{1} \\
\eta^{+}_{2}          
\end{array} \right) 
\sim ({\bf 1},{\bf 3},0),\quad
\rho = 
\left( \begin{array}{c} 
\rho^{+} \\ 
\rho^{0} \\
\rho^{++}          
\end{array} \right) 
\sim ({\bf 1},{\bf 3},+1), \,\
\chi = 
\left( \begin{array}{c} 
\chi^{-} \\ 
\chi^{--} \\
\chi^{0}          
\end{array} \right) 
\sim ({\bf 1},{\bf 3},-1), \nonumber \\
S &=& \left( \begin{array}{ccc} 
\sigma^{0}_{1}& 
\frac{h^{+}_{2}}{ \sqrt{2}}& \frac{h^{-}_{1}}{ \sqrt{2}} \\ 
\frac{h^{+}_{2}}{ \sqrt{2}}& H^{++}_{1}& \frac{ \sigma^{0}_{2}}{ \sqrt{2}} \\
 \frac{h^{-}_{1}}{ \sqrt{2}}& 
\frac{\sigma^{0}_{2}}{ \sqrt{2}}&  H^{--}_{2}        
\end{array} \right) \sim ({\bf1},{\bf \bar{6}},0),
\end{eqnarray}
We associate each scalar field a chiral or anti-chiral superfield given by
\begin{equation}
\Phi(y, \theta)=A(y)+ \theta \tilde{A}(y)+ \theta \theta F_{A}(y),
\end{equation}
where $A$ represents a scalar field while $\tilde{A}$ is its 
superpartner known as higgsinos.

The higgsinos $\tilde{\eta}$, $\tilde{\rho}$, $\tilde{\chi}$ and $\tilde{S}$, 
presented above, generate chiral anomalies and to cancel them we have 
to introduce the followings scalars
\begin{eqnarray}
\eta^{\prime} &=& 
\left( 
\begin{array}{c} 
\eta^{\prime0} \\ 
\eta^{\prime+}_{1} \\
\eta^{\prime-}_{2}          
\end{array} \right) 
\sim ({\bf1},{\bf \bar{3}},0),\quad
\rho^{\prime} = 
\left( \begin{array}{c} 
\rho^{\prime-} \\ 
\rho^{\prime0} \\
\rho^{\prime--}          
\end{array} 
\right) 
\sim ({\bf1},{\bf \bar{3}},-1), \,\   
\chi^{\prime} = 
\left( \begin{array}{c} 
\chi^{\prime+} \\ 
\chi^{\prime++} \\
\chi^{\prime0}          
\end{array} \right) 
\sim ({\bf1},{\bf \bar{3}},+1) \nonumber \\
S^{\prime} &=& 
\left( \begin{array}{ccc} 
\sigma^{\prime0}_{1}& \frac{h^{\prime-}_{2}}{ \sqrt{2}}& 
\frac{h^{\prime+}_{1}}{ \sqrt{2}} \\ 
\frac{h^{\prime-}_{2}}{ \sqrt{2}}&H^{\prime--}_{1}& 
\frac{ \sigma^{ \prime 0}_{2}}{ \sqrt{2}} \\
 \frac{h^{\prime+}_{1}}{ \sqrt{2}}& 
\frac{ \sigma^{ \prime 0}_{2}}{ \sqrt{2}}&  
H^{\prime ++}_{2}        
\end{array} \right) \sim ({\bf1},{\bf6},0).
\label{escoriginais} 
\end{eqnarray}

In the MSUSY331 we need to introduce the following three vector superfields $\hat{V}^{a}_{C}\sim({\bf 8},{\bf 1}, 0)$
\footnote{The gluinos are the superpartner of gluons, and therefore they are in the adjoint representation of $SU(3)$, wchich is real.}, 
where $a=1,2, \ldots ,8$, $\hat{V}^{a}\sim({\bf 1},{\bf 8}, 0)$, and 
$\hat{V}\sim({\bf 1},{\bf 1}, 0)$. The particle content 
in each vector superfield is presented in the Tab.(\ref{gaugemssm}). The vector superfield
\footnote{It is also known as Real Superfield due the fact that this superfield satisfy the following constraint 
$V(x, \theta, \bar{ \theta})=V^{\dagger}(x, \theta, \bar{ \theta})$.} 
in the Wess-Zumino gauge is written as 
\begin{equation}
V_{WZ}(x, \theta, \bar{ \theta}) = - \left( \theta \sigma^{m} \bar{ \theta}\right) v_{m}(x) +
\imath \left( \theta \theta \right) \left( \bar{ \theta} \bar{ \lambda}(x) \right) 
- \imath \left( \bar{ \theta} \bar{ \theta} \right) \left( \theta \lambda(x) \right)
+ \frac{1}{2} \left( \theta \theta \right) \left( \bar{ \theta} \bar{ \theta} \right) D(x).
\label{eq:vwzdef}
\end{equation}
$v_{m}$ is a gauge boson, $\lambda$ is their superpartner known as gauginos. For 
people interested in the Lagrangian of this model, see the 
references \cite{331susy1,mcr}.
\begin{table}[h]
\begin{center}
\begin{tabular}{|c|c|c|c|}
\hline 
${\rm{Vector \,\ Superfield}}$ & ${\rm{Gauge \,\ Bosons}}$ & ${\rm{Gaugino}}$ & ${\rm Gauge \,\ constant}$ \\
\hline 
$\hat{V}^{a}_{C}\sim({\bf 8},{\bf 1}, 0)$ & $g^{a}_{m}$ & $\lambda _{C}^{a}$ & $g_{s}$ \\
\hline 
$\hat{V}^{a}\sim({\bf 1},{\bf 8}, 0)$ & $V^{a}_{m}$ & $\lambda^{a}$ & $g$ \\
\hline
$\hat{V}\sim({\bf 1},{\bf 1}, 0)$ & $V_{m}$ & $\lambda$ & $g^{\prime}$ \\
\hline
\end{tabular}
\end{center}
\caption{\small Particle content in the vector superfields in MSUSY331.}
\label{gaugemssm}
\end{table}

The superpotential of our model is given by
\begin{equation}
W=W_{2}+W_{3}+hc, 
\label{sp1}
\end{equation}
with $W_{2}$ has only two chiral superfields while $W_{3}$ has three chiral superfields. The terms allowed by  
our symmetry are
\begin{eqnarray}
W_{2}&=&\sum_{i=1}^{3}
\mu_{0i}(\hat{L}_{i}\hat{\eta}^{\prime})
+ 
\mu_{ \eta} 
(\hat{\eta}\hat{\eta}^{\prime})
+
\mu_{ \rho} 
(\hat{\rho}\hat{\rho}^{\prime})
+ 
\mu_{ \chi} 
(\hat{\chi}\hat{\chi}^{\prime})
+
\mu_{S} Tr[(\hat{S} \hat{S}^{\prime})], 
\label{w2geral}
\end{eqnarray}
All the coefficients in the equation above 
have mass dimension and we also have
\begin{eqnarray}
W_{3}&=&\sum_{i=1}^{3} \left[ \sum_{j=1}^{3} \left( 
\sum_{k=1}^{3}\lambda_{1ijk} 
(\epsilon \hat{L}_{i}\hat{L}_{j}\hat{L}_{k})+
\lambda_{2ij} (\epsilon \hat{L}_{i}\hat{L}_{j}\hat{\eta})+
\lambda_{3ij} 
(\hat{L}_{i}\hat{S}\hat{L}_{j}) \right)+
\lambda_{4i}(\epsilon\hat{L}_{i}\hat{\chi}\hat{\rho})
\right] \nonumber \\
&+&
f_{1} 
(\epsilon \hat{\rho} \hat{\chi}\hat{\eta})
+
f_{2} 
(\hat{\eta}\hat{S}\hat{\eta})
+
f_{3} 
(\hat{\chi}\hat{S}\hat{\rho})  
+
f_{4}\epsilon_{ijk}\epsilon_{lmn}\hat{S}_{il}
\hat{S}_{jm}\hat{S}_{kn}+
f^{\prime}_{1} (\epsilon \hat{\rho}^{\prime}\hat{\chi}^{\prime} \hat{\eta}^{\prime})+
f^{\prime}_{2} (\hat{\eta}^{\prime} 
\hat{S}^{\prime}\hat{\eta}^{\prime})  
\nonumber \\ &+&
f^{\prime}_{3} (\hat{\chi}^{\prime} 
\hat{S}^{\prime}\hat{\rho}^{\prime}) 
+
f^{\prime}_{4}\epsilon_{ijk}\epsilon_{lmn}
\hat{S}^{\prime}_{il}\hat{S}^{\prime}_{jm}
\hat{S}^{\prime}_{kn} +
\sum_{i=1}^{3} \left[ \kappa_{1i}
(\hat{Q}_{1} \hat{\eta}^{\prime}) \hat{u}^{c}_{i}+ 
\kappa_{2i} 
(\hat{Q}_{1} \hat{\rho}^{\prime}) \hat{d}^{c}_{i} \right]
+
\kappa_{3} 
(\hat{Q}_{1}\hat{\chi}^{\prime}) 
\hat{J}^{c} 
\nonumber \\
&+&
\sum_{\alpha =2}^{3} \left[
\sum_{i=1}^{3} \left(
\kappa_{4\alpha i} 
(\hat{Q}_{\alpha} \hat{\eta}) \hat{d}^{c}_{i}+
\kappa_{5\alpha i} 
(\hat{Q}_{\alpha} \hat{\rho}) \hat{u}^{c}_{i} \right) 
+
\sum_{\beta =1}^{2} 
\kappa_{6\alpha\beta}
(\hat{Q}_{\alpha} \hat{\chi}) \hat{j}^{c}_{\beta} 
+ \sum_{i=1}^{3}\sum_{j=1}^{3}
\kappa_{7\alpha ij} 
(\hat{Q}_{\alpha} \hat{L}_{i}) 
\hat{d}^{c}_{j} \right]
\nonumber \\
&+&
\sum_{i=1}^{3} \left\{ \sum_{j=1}^{3}\left[ \sum_{k=1}^{3}
\xi_{1ijk} \hat{d}^{c}_{i} \hat{d}^{c}_{j} \hat{u}^{c}_{k}
+
\sum_{\beta =1}^{2} \left(
\xi_{2ij\beta} \hat{u}^{c}_{i} \hat{u}^{c}_{j} \hat{j}^{c}_{\beta}
\right]+
\xi_{3 i\beta} \hat{d}^{c}_{i} \hat{J}^{c} \hat{j}^{c}_{\beta} \right) 
\right\}. 
\label{sp3m1}
\end{eqnarray}
All the coefficients in $W_{3}$ are dimensionless. Note that the scalar 
$S^{\prime}$ does not have any type of coupling with the fermions of the model, so the VEV of its neutral components can be a few GeV, for more 
details about the masses of fermions see reference \cite{331susy1}.

We want to draw your attention to the following facts
\begin{itemize}
\item A familly of quarks has different transformation properties than the 
others two families, see our Eqs.(\ref{q1l},\ref{q23l});
\item We need two different VEV to generate masses for the up quarks, 
$\eta^{\prime}$ and $\rho$, and down quarks, $\rho^{\prime}$ and $\eta$, 
as discussed briefly in our reference \cite{331susy1};
\end{itemize} 
with this we can hope to have a simple explanation of the why the quarks of 
the third family are much more massive than the quarks of the other two 
families in a mechanism similar as presented by references \cite{banks,cmmc,cmmc1}. 

If we put the following discrete symmetry
\begin{equation}
\hat{u}^{c}_{3} \rightarrow - \hat{u}^{c}_{3}, \,\ 
\hat{d}^{c}_{3} \rightarrow - \hat{d}^{c}_{3},
\end{equation}
we obtain the following expressions for the masses of the quarks 
\cite{Rodriguez:2010tn,Montero:2001ch}
\begin{eqnarray}
m_{t}= \kappa _{13}v_{\eta^{\prime}}, \,\ 
m_{b}= \kappa _{23}v_{\rho^{\prime}},
\end{eqnarray} 
the masses of the others quarks are
\begin{eqnarray}
X_{u}&\simeq&\left( 
\begin{array}{cc}
\kappa_{511} & \kappa_{521}  \\
\kappa_{512} & \kappa_{522}  
\end{array}
\right) v_{\rho}, \,\
X_{d}\simeq \left( 
\begin{array}{cc}
\kappa_{411} & \kappa_{412}  \\ 
\kappa_{421} & \kappa_{422} 
\end{array}
\right) v_{\eta}.
\label{massaproxx}
\end{eqnarray}
We intend to present a much more detailed numerical study, using 
part of this article, regarding the masses of all fermions and show we 
can fit in this model the current measurement of the top quark is \cite{Sirunyan:2018gqx}  
\begin{equation}
m_{t}=172.25\pm 0.08\mathrm{(stat.)} \pm 0.62\mathrm{(syst.)}~\mathrm{GeV},
\label{mtexp}
\end{equation}
as well the masses of other quarks in MeV \cite{pdg} 
\begin{equation}
m_{u}\sim1-5,\,\ m_{d}\sim3-9,\,\ m_{s}\sim75-170,
\end{equation}
while the masses of charm and bottom mass in GeV \cite{pdg}
\begin{equation}
m_{c}\sim 1.15-1.35, \,\ m_{b}=4.4.
\end{equation}

Recently it was reported a light Higgs boson at the LHC and its value is given by \cite{Aad:2015zhl}
\begin{equation}
M_{H} = 125.09 \pm 0.21\,\mathrm{(stat.)} \pm 0.11\,\mathrm{(syst.)}~\mathrm{GeV}.
\end{equation}
We have shown the lighest higgs has mass around \cite{331susy1,esc2} 
\begin{equation}
M_{H} \simeq 125 GeV ~\mathrm{GeV}.
\label{higgsmassMSUSY331}
\end{equation}

The charged gauge bosons of this model are \cite{mcr}
\begin{eqnarray}
W^{ \pm}_{m}(x)&=&-\frac{1}{\sqrt{2}}(V^{1}_{m}(x) 
\mp i V^{2}_{m}(x)), \,\
V^{ \pm}_{m}(x)=-\frac{1}{\sqrt{2}}(V^{4}_{m}(x) 
\pm i V^{5}_{m}(x)) \nonumber \\
U^{\pm \pm}_{m}(x) &=&- \frac{1}{\sqrt{2}}(V^{6}_{m}(x) 
\pm i V^{7}_{m}(x)).
\label{defcarbosons}
\end{eqnarray}
The interaction between the charged bosons with the fermions are given by \cite{Rodriguez:2010tn}
\begin{eqnarray}
{\cal L}_l^{CC}&=&-\frac{g}{\sqrt{2}}\sum_{l}\left(\bar{\nu}_{lL}\gamma^{m}V_{\rm PMNS}l_{L}W^{+}_{m}+ 
\bar{l}^{c}_{L}\gamma^{m}U_{V}\nu_{lL} V^{+}_{m}+
\bar{l}^{c}_{L}\gamma^{m} U_{U}l_{L}U^{++}_{m}+hc \right), \nonumber \\
{\cal L}_q^{CC}& = & -\frac{g}{2\sqrt{2}}
\left[\overline{U}\gamma^{m}(1 - \gamma_{5})V_{\rm CKM}DW^{+}_{m}  + 
\overline{U}\gamma^{m}(1 - \gamma_{5})\zeta{\cal JV}_m + 
\overline{D}\gamma^{m}(1 - \gamma_{5})\xi{\cal JU}_m\right] \nonumber \\
&+& hc, 
\nonumber \\
\label{lq}
\end{eqnarray}
where we have defined the mass eigenstates in the following way
\begin{eqnarray}
U = \left(\begin{array}{c} u \\ c \\ t
\end{array}\right), \quad
D & = & \left(\begin{array}{c}
         d \\ s \\ b
\end{array}\right), \quad
{\cal V}_m = \left(\begin{array}{c}
         V^+_m \\ U^{--}_m \\ U^{--}_m\end{array}\right), \nonumber \\
{\cal U}_m & = & \left(\begin{array}{c}
         U^{--}_m \\ V^+_m \\ V^+_m
\end{array}\right), 
\end{eqnarray}\label{maest}
and ${\cal J} = {\rm diag}\left(\begin{array}{ccc}J_1 & J_2 & 
J_3\end{array}\right)$. The $V_{\rm PMNS}$ is the Pontecorvo-Maki-Nakagawa-Sakata mixing matrix and $V_{\rm CKM}$ is the usual 
Cabibbo-Kobayashi-Maskawa mixing matrix, see comments after Eq.(\ref{yukawaquarks}). There are new mixing matrices given by $U_{V}$, $U_{U}$, $\xi$ 
and $\zeta$ containing new unknown mixing parameters.

We have already showed that in the M\o ller scattering and in 
muon-muon scattering we can show that left-right asymmetries $A_{RL}(ll)$  are very sensitive to a 
doubly charged vector bilepton resonance but they are insensitive to scalar ones 
\cite{Montero:1998ve,Montero:1999en,Montero:2000ch}. 

The neutral gauge bosons are the photon
\begin{eqnarray}
A_{m}&=& \sin \theta_{W} \left( V^{3}_{m}- \sqrt{3}V^{8}_{m} \right) + 
\sqrt{1-4 \sin^{2} \theta_{W}}V_{m},
\end{eqnarray}
where $\theta_{W}$ is the Weinberg angle and it is defined by the following 
algebric relation
\begin{equation}
\frac{M^{2}_{Z}}{M^{2}_{W}}\equiv 
\frac{1}{\cos^{2} \theta_{W}}= 
\frac{1}{1- \sin^{2} \theta_{W}},
\label{tinthetaw}
\end{equation} 
and we have also two massives neutral bosons defined as
\begin{eqnarray}
Z_{m}&=& \cos \theta_{W}V^{3}_{m}+ \sqrt{3}\tan \theta_{W}\sin \theta_{W}V^{8}_{m}- 
\tan \theta_{W}\sqrt{1-4 \sin^{2} \theta_{W}}V_{m}, \nonumber \\
Z^{\prime}_{m}&=& \frac{1}{\cos \theta_{W}}\left[
\sqrt{1-4 \sin^{2} \theta_{W}}V^{8}_{m}+ \sqrt{3}\sin \theta_{W}V_{m} \right].
\end{eqnarray}

Similarly, we have the neutral currents coupled to both $Z^0$ and
$Z'^0$ massive vector bosons, according to the Lagrangian
\begin{equation}
{\cal L}_\nu^{NC}=-\frac{g}{2}\frac{M_Z}{M_W}
\bar\nu_{lL}\gamma^m\nu_{lL}
\left[ Z_m-\frac{1}{\sqrt3}\frac{1}{\sqrt{h(t)}}Z^{\prime}_m \right] ,
\label{e27}
\end{equation}
with $h(t)=1+4t^2$, for neutrinos and
\begin{equation}
{\cal L}_l^{NC} =-\frac{g}{4}\frac{M_Z}{M_W} \left[ \bar
l\gamma^m(v_l+a_l\gamma^5)lZ_m+ \bar
l\gamma^m(v'_l+a'_l\gamma^5)lZ^{\prime}_m \right] ,
\label{e28}
\end{equation}
for the charged leptons, where we have defined
\begin{eqnarray}
\begin{array}{clccc}
v_l= & -1/h(t),&a_l=& 1,& \nonumber \\
v'_l=& -\sqrt{3/h(t)},
& a'_l=& v'_l/3.& \nonumber
\end{array}
\nonumber
\end{eqnarray}
We can use muon collider to discover the new neutral $Z^{\prime}$ boson using the reaction 
$\mu e \to \mu e$ it was shown at \cite{Montero:1999en,Montero:1998sv} that $A_{RL}(\mu e)$ asymmetry is
considerably enhanced.

Before we continue presenting 
MSUSY331, we would like to highlight that those new exotic quarks, $J$, $j_{1}$ 
and $j_{2}$, may be discovered by the Large Hadron Collider (LHC) thought 
$pp$ collisions, via the following subprocess\footnote{I want to thank Alexander S. Belyaev, who brought this process to my attention at the 
end of my PhD studies at IFT-Unesp, but unfortunately we were unable to publish this study together.}
\begin{eqnarray}
g+d \to U^{--}+J, \,\
g+u \to  U^{--}+j_{\alpha}, 
\label{intersting}
\end{eqnarray} 
its signature is $llXX$ and it can be descted at LHC if they really exist in 
nature \cite{Rodriguez:2010tn,dutta}.

In order to get the gauge bosons masses we have to calculate
\begin{eqnarray}
{\cal L}_{\rm Higgs}&=& ({\cal D}_{m} \eta)^{\dagger}({\cal D}^{m} \eta)+ 
({\cal D}_{m} \rho)^{\dagger}({\cal D}^{m} \rho)+
({\cal D}_{m} \chi)^{\dagger}({\cal D}^{m} \chi)+
(\overline{{\cal D}_{m}} \eta^{\prime})^{\dagger}(\overline{{\cal D}^{m}} \eta^{\prime}) \nonumber \\ &+&
(\overline{{\cal D}_{m}} \rho^{\prime})^{\dagger}(\overline{{\cal D}^{m}} \rho^{\prime})+
(\overline{{\cal D}_{m}} \chi^{\prime})^{\dagger}(\overline{{\cal D}^{m}} \chi^{\prime})+
Tr[({\cal D}_{m}S)^{\dagger}({\cal D}^{m}S)] \nonumber \\
&+&
Tr[(\overline{{\cal D}_{m}}S^{\prime})^{\dagger}(\overline{{\cal D}^{m}}S^{\prime})]. \nonumber \\
\label{lagHHVV}
\end{eqnarray}

The  mass of charged gauge boson are given by\footnote{About 
$W$-mass, see our Eq.(\ref{masswezgelmini}), and remember 
that $v_{\sigma_{1}}$ belongs to a triplet of $SU(2)$ while 
$v_{\sigma_{2}}$ for doublet under this same group, see Eq.(\ref{sig1:triplet}).}  
\begin{eqnarray}
M^{2}_{W}&=& \frac{g^{2}}{4} \left[
v^{2}_{ \eta}+v^{2}_{ \rho}+v^{2}_{ \sigma_{2}}+v^{2}_{ \eta^{\prime}}+v^{2}_{ \rho^{\prime}} +
v^{2}_{\sigma^{\prime}_{2}} +
2 \left( v^{2}_{\sigma_{1}}+v^{2}_{\sigma^{\prime}_{1}} \right)  \right], \nonumber \\
M^{2}_{V}&=& \frac{g^{2}}{4}\left[
v^{2}_{ \eta}+v^{2}_{ \chi}+v^{2}_{ \sigma_{2}}+v^{2}_{ \eta^{\prime}}+v^{2}_{ \chi^{\prime}}+
v^{2}_{\sigma^{\prime}_{2}}+ 
2 \left( v^{2}_{\sigma_{1}}+v^{2}_{\sigma^{\prime}_{1}} \right)  \right], \nonumber \\
\delta_{WV}&=&\frac{g^{2}}{\sqrt{2}}\left(
v_{\sigma_{1}}v_{\sigma_{2}}+ 
v_{\sigma^{\prime}_{1}}v_{\sigma^{\prime}_{2}} 
\right), \nonumber \\
M^{2}_{U}&=& \frac{g^{2}}{4} \left(
v^{2}_{\rho}+v^{2}_{\chi}+4v^{2}_{\sigma_{2}}+
v^{2}_{\rho^{\prime}}+v^{2}_{\chi^{\prime}}+
4v^{2}_{\sigma^{\prime}_{2}} 
\right). 
\label{bgmcc}
\end{eqnarray}
This new charged vector boson, $V^{\pm}$, is best known 
in the literature as being $W^{\prime}$ and in recent 
analyzes presented by the LHC's ATLAS and CMS collaborations 
have established the following limit for its mass \cite{pdg} 
\begin{equation}
0.15 \leq M_{W^{\prime}} \leq 7 \,\ {\mbox TeV},
\label{explimV}
\end{equation}
there are also the following High-Precision limit on its mass \cite{Pankov:2019yzr,Osland:2020onj} 
\begin{equation}
M_{W^{\prime}} > 6 \,\ {\mbox TeV},
\label{explimV1}
\end{equation}

The single charged gauge bosons can mixing and we present the 
analytical analysis of this mixture in 
App.(\ref{sec:mixingchar}). It is possible the following lower limit 
on the mass of $U$-boson \cite{barela,barela1}
\begin{equation}
M_{U}\geq 200 \,\ {\mbox GeV}.
\label{lowerlimitU}
\end{equation}

We also have, as in SM, the usual $Z$-boson and an extra 
neutral gauge boson known as $Z^{\prime}$-boson and in the 
approximation that 
$v_{ \chi} \simeq v_{ \chi^{\prime}}$ 
together with the extra assumption that these two VEVs are 
the largest in this model, allow us to write the following 
approximate results \cite{pp,mcr}
\begin{eqnarray}
M^{2}_{Z}&\approx& \frac{1}{2} 
\left( \frac{g^{2}+4g^{\prime 2}}{g^{2}+3g^{\prime 2}} \right)
\left[ v^{2}_{ \eta}+v^{2}_{ \rho}+v^{2}_{ \sigma_{2}}+v^{2}_{ \eta^{\prime}}+v^{2}_{ \rho^{\prime}}+
v^{2}_{\sigma^{\prime}_{2}}+ 
4 \left( v^{2}_{\sigma_{1}}+v^{2}_{\sigma^{\prime}_{1}} \right)  \right], \nonumber \\
M^{2}_{Z^{\prime}}&\approx& \frac{2}{3} (g^{2}+3g^{\prime 2}) 
(v^{2}_{ \chi}+v^{2}_{ \chi^{\prime}}),
\label{zmassmcr}
\end{eqnarray}
The neutral gauge bosons can mixing and we present the analytical analysis 
of this mixture in App.(\ref{sec:mixingneu}) and
the experimental limits on the mass of the new heavy gauge 
boson, $Z^{\prime}$, is as follows \cite{pdg} 
\begin{equation}
M_{Z^{\prime}} > 1.1 \,\ {\mbox TeV},
\end{equation}
there are also the following High-Precision limit on its mass 
\cite{Pankov:2019yzr,Osland:2020onj} 
\begin{equation}
M_{Z^{\prime}} > 5 \,\ {\mbox TeV}.
\label{explimZP}
\end{equation}  

Using the first expression from Eq.(\ref{bgmcc}), together with $M^{2}_{Z}$, we can write\footnote{We are considering, at this point only, 
$v_{\sigma_{1}}=v_{\sigma^{\prime}_{1}}=0$ as presented on the previous study, see reference \cite{pp}.}
\begin{equation}
\frac{M^{2}_{Z}}{M^{2}_{W}}= \frac{1+4t^{2}}{1+3t^{2}}, \,\ 
t \equiv \frac{g^{\prime}}{g}.
\label{tinthetawm331}
\end{equation}
When we impose Eq.(\ref{tinthetaw}) is equal to Eq.(\ref{tinthetawm331}), 
we obtain the famous relationship
\begin{equation}
t^{2}= \frac{\sin^{2} \theta_{W}}{1-4 \sin^{2} \theta_{W}},
\label{tinthetaw2}
\end{equation}
To avoid the loss of the model being perturbative, it allows us to write the following inequality
\begin{eqnarray}
\sin^{2}\theta_{W}< \frac{1}{4},
\label{muesc}
\end{eqnarray} 
when we have equality, we obtain the Landau pole of this model, that is, to obtain this pole, we impose
\begin{eqnarray}
\sin^{2}\theta_{W}(\mu)= \frac{1}{4},
\end{eqnarray}
the Landau scale value of the model without and with SUSY, 
respectively, has the following values \cite{Dias:2004dc}
\begin{eqnarray}
\mu_{\rm M331}&=&5.7 \,\ {\rm TeV}, \nonumber \\
\mu_{\rm MSUSY331}&=&7.8 \,\ {\rm TeV},
\end{eqnarray}
recently a new analysis appeared in which we obtain the following value 
\cite{Barela:2023oyp}
\begin{eqnarray}
\mu_{\rm M331}&=&8.5 \,\ {\rm TeV}.
\label{barelalimite}
\end{eqnarray}

\section{Numerical Analyses for CDF Experimental Results.}
\label{wcdf}

In a previous work, we have choose the following VEVs for 
our scalars (in GeV) \cite{331susy1}
\begin{equation}
v_{\eta}=20, \,\ v_{\sigma_{2}}=10, \,\ 
v_{\eta^{\prime}}=v_{\rho^{\prime}}=1,
\label{standardvevm331}
\end{equation}
then we fix the values of $v_{\rho}$ to explain the mass values
for $W$-mass at SM and we get the following value
\begin{equation}
v_{\rho}=245.198 \,\ {\rm GeV},
\label{fixrho}
\end{equation}
in this way we also fix the numerical value of the mass of 
$Z$, using Eq.(\ref{tinthetaw}), however, we will return to 
this subject later in our Eqs.(\ref{mzhip1},\ref{ZSM}).

\begin{figure}[ht]
\begin{center}
\vglue -0.009cm
\mbox{\epsfig{file=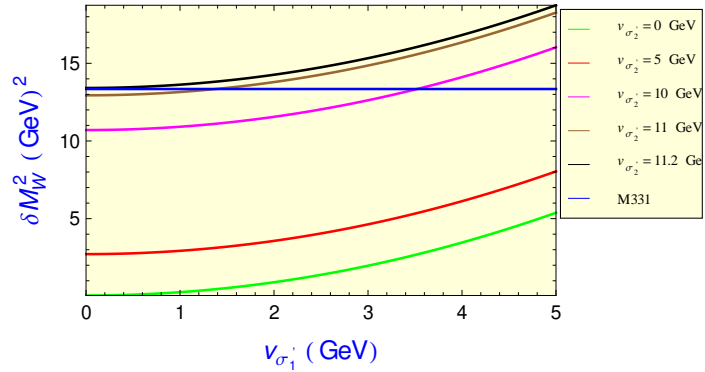,width=0.7\textwidth,angle=0}}       
\end{center}
\caption{$\delta M^{2}_{W}$, defined in Eq.(\ref{resmsusy331}), 
as function of $v_{\sigma^{\prime}_{1}}$, for some values ​​of 
$v_{\sigma^{\prime}_{2}}$ shown in the square while 
M331 is the numerical defined in our Eq.(\ref{expvalmassdiff}).}
\label{deltawvvs1p}
\end{figure}

\begin{figure}[ht]
\begin{center}
\vglue -0.009cm
\mbox{\epsfig{file=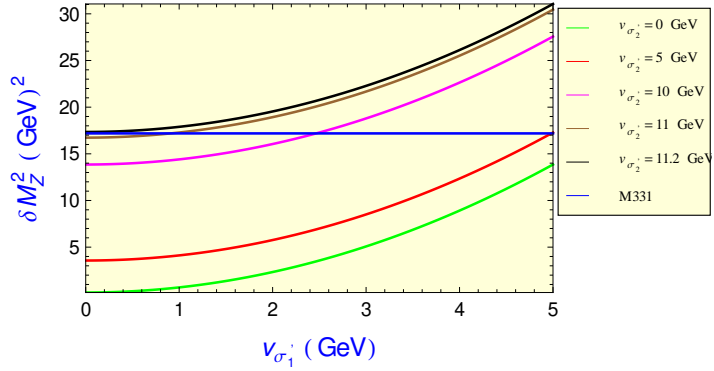,width=0.7\textwidth,angle=0}}       
\end{center}
\caption{$\delta M^{2}_{Z}$, defined in Eq.(\ref{resmsusy331}), 
as function of $v_{\sigma^{\prime}_{1}}$, for some values ​​of 
$v_{\sigma^{\prime}_{2}}$ shown in the square while 
M331 is defined in our Eq.(\ref{dmzdmwm331}).}
\label{deltazvvs1p}
\end{figure}

The masses of the $W$-boson and $Z$-boson receive 
a new tree-level contribution\footnote{We used 
Eqs.(\ref{expvalmassdiff},\ref{m331anomalia}).}, which is given by
\begin{eqnarray}
\delta M^{2}_{W}&=& \frac{g^{2}}{4}\left[ v^{2}_{\sigma^{\prime}_{2}} +2 \left(
v^{2}_{\sigma^{1}}+v^{2}_{\sigma^{\prime}_{1}} \right) 
 \right], \,\
\delta M^{2}_{Z}= \frac{g^{2}}{4 \cos^{2}\theta_{W}} \left[ v^{2}_{\sigma^{\prime}_{2}} + 4 \left(
v^{2}_{\sigma^{1}}+v^{2}_{\sigma^{\prime}_{1}} \right)  \right], \nonumber \\
\label{resmsusy331}
\end{eqnarray}
therefore, as first prevision of this model is
\begin{equation}
\delta M^{2}_{W} \neq \delta M^{2}_{Z},
\end{equation}
our analytical results are in agreement with the 
results obtained for the M331 as presented by the reference 
\cite{Pisano:1997hk} and also in our 
Eqs.(\ref{dmzdmwm331},\ref{mzcdfvaluesm331}). Before starting 
our numerical analysis\footnote{We are also using the others VEVs given at 
Eqs.(\ref{standardvevm331},\ref{fixrho})}, we will present the graphs of the 
variables 
in Eq.(\ref{resmsusy331}), in terms of $v_{\sigma^{\prime}_{1}}$ 
with $v_{\sigma_{1}}=0.5$ GeV for some values ​​of $v_{\sigma^{\prime}_{2}}$, in 
Figs.(\ref{deltawvvs1p},\ref{deltazvvs1p}) respectively. We realize 
that we can explain both variations considering 
$10<v_{\sigma^{\prime}_{2}}<11.2$ GeV and with $0<v_{\sigma^{\prime}_{1}}<4$ GeV, in other words, as we wish, using 
the VEVs from the triplets having values ​​on the order of a few GeV.


To explain the mass anomaly $W$, defined by 
Eq.(\ref{expvalmassdiff}), we have to impose
\begin{equation}
\left| \sqrt{2 \left( v^{2}_{\sigma_{1}}+v^{2}_{\sigma^{\prime}_{1}}\right) +
v^{2}_{\sigma^{\prime}_{2}}} \right|= 
11.19 \,\ {\rm GeV}.
\label{wmassrest}
\end{equation}
All VEV that enter this equation are extra fields in 
relation to M331, do not have any restrictions arising from 
the fermion mass explanation, see our Eq.(\ref{sp3m1}).

By another hand, for the $\rho$-parameter in our model, see our Eq.(\ref{rhoparametergelmini}), must satisfy the 
following algebraic relationship
\begin{eqnarray}
\rho &=& \frac{1+R}{1+R^{\prime}} \nonumber \\
R&=&\frac{v^{2}_{\sigma^{\prime}_{2}}}{v^{2}_{\rm SM}}+ 2 \left( \frac{v^{2}_{\sigma_{1}}+v^{2}_{\sigma^{\prime}_{1}}}{v^{2}_{\rm SM}} 
\right), \,\
R^{\prime}=\frac{v^{2}_{\sigma^{\prime}_{2}}}{v^{2}_{\rm SM}}+ 
4 \left( 
\frac{v^{2}_{\sigma_{1}}+v^{2}_{\sigma^{\prime}_{1}}}{v^{2}_{\rm SM}} 
\right)  \nonumber \\
v^{2}_{\rm SM}&=&v^{2}_{ \eta}+v^{2}_{ \rho}+v^{2}_{ \sigma_{2}}+
v^{2}_{ \eta^{\prime}}+v^{2}_{ \rho^{\prime}} \nonumber \\
\rho &\simeq& 1+R-R^{\prime}=1-2 
\left( 
\frac{v^{2}_{\sigma_{1}}+v^{2}_{\sigma^{\prime}_{1}}}{v^{2}_{\rm SM}} 
\right).
\label{rhoexpmsusy331}
\end{eqnarray} 
We fix the following energy scale for the scalars belonging to the triplet of the subgroup of $SU(2)$
\begin{eqnarray}
\sqrt{ v^{2}_{\sigma_{1}}+v^{2}_{\sigma^{\prime}_{1}}} < 
\sqrt{\frac{0.0008}{2}}v_{\rm SM} \approx 5.5 \,\ {\rm GeV}.
\label{rhorest}
\end{eqnarray}

Therefore, we can explain at the same time the new data on 
the mass of the W boson presented by the CDF collaboration, 
Eq.(\ref{wmassrest}), as well as the parameter $\rho$,
Eq.(\ref{rhorest}), if the new VEV, coming from the $SU(2)$ 
triplets, are smaller then 5.5 GeV. Next we will explore 
three hypotheses.

We will define the first hypothesis, as being where all the 
VEV of the triplet and anti-triplet are zero, this means
\begin{equation}
v_{\sigma_{1}}=v_{\sigma^{\prime}_{1}}=0 \,\ {\rm GeV,}
\label{hip1}
\end{equation}
in other words, we recover the SM, as only fields in the 
$SU(2)$ doublet representation acquire non-zero VEV.

\begin{figure}[ht]
\begin{center}
\vglue -0.009cm
\mbox{\epsfig{file=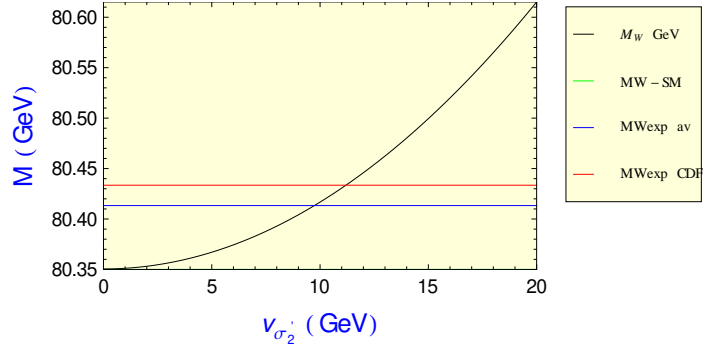,width=0.7\textwidth,angle=0}}       
\end{center}
\caption{The masses of $W$-boson, others VEVs are shown at 
Eqs.(\ref{standardvevm331},\ref{fixrho}), as function of $v_{\sigma^{\prime}_{2}}$, the green line is given by 
Eq.(\ref{smvalue}), while the blune line is the average value of this parameter, Eq.(\ref{average}), and red 
line show the results of CDF presented in 
Eq.(\ref{cdfresult}).}
\label{mwvvs2p}
\end{figure}

Our first results, in this hypothesis, are shown in
Fig.(\ref{mwvvs2p}), where we see that we can explain the 
average value of experimental measurements of the mass of the 
$W$-boson if $v_{\sigma^{\prime}_{2}}\sim 9.74$ GeV
and the CDF result when $v_{\sigma^{\prime}_{2}}\sim 11.19$ 
GeV. 

\begin{figure}[ht]
\begin{center}
\vglue -0.009cm
\mbox{\epsfig{file=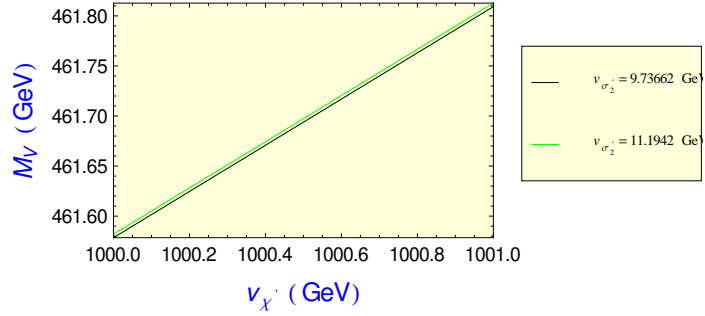,width=0.7\textwidth,angle=0}}       
\end{center}
\caption{The masses of $V$-boson as function of $v_{\chi^{\prime}}$ and $v_{\chi}=1000$ GeV, the black line is for $v_{\sigma^{\prime}_{2}}\sim 9.74$ GeV, while the green line we use $v_{\sigma^{\prime}_{2}}\sim 11.19$ GeV, 
the others VEVs are shown at 
Eqs.(\ref{standardvevm331},\ref{fixrho},\ref{hip1}) and the masses are in agreement 
with the experimental bound given in Eq.(\ref{explimV}), but these values 
​​do not satisfy the constraint given by Eq.(\ref{explimV1}), we will show in 
below that we can satisfy this link coming from High-precision 
limit.}
\label{mvvchi}
\end{figure}

\begin{figure}[ht]
\begin{center}
\vglue -0.009cm
\mbox{\epsfig{file=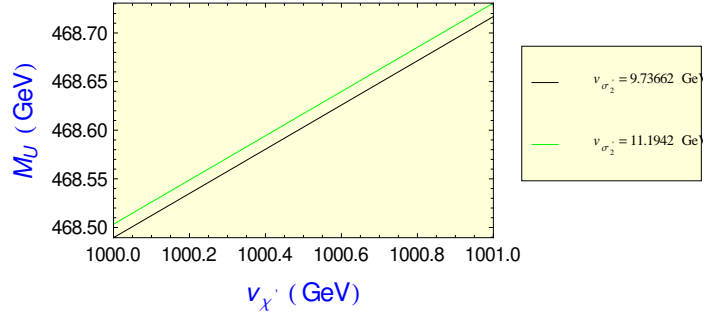,width=0.7\textwidth,angle=0}}       
\end{center}
\caption{The masses of $U$-boson as function of $v_{\chi^{\prime}}$ and $v_{\chi}=1000$ GeV, the black line is for $v_{\sigma^{\prime}_{2}}\sim 9.74$ GeV, while the green line we use $v_{\sigma^{\prime}_{2}}\sim 11.19$ GeV, 
the others VEVs are shown at Eqs.(\ref{standardvevm331},\ref{fixrho},\ref{hip1}) and it is in agreement with Eq.(\ref{lowerlimitU}).}
\label{muvchi}
\end{figure}

In this case there is no mixing in the single charged 
gauge bosons, see our Eq.(\ref{bgmcc}). We show the masses of 
$M_{V}$ and $M_{U}$ in our Figs.(\ref{mvvchi},\ref{muvchi}). 
For both new gauge bosons we obtain higher mass values ​​for 
$v_{\sigma^{\prime}_{2}}\sim 11.19$ GeV in relation to 
$v_{\sigma^{\prime}_{2} }\sim 9.73$ GeV and the differences 
between the values ​​are greater for $U^{\pm \pm}$-boson than 
for the $V^{\pm}$-bosons. We can also conclude $M_{U}>M_{V}$, 
this result is in agreement with the previous analysis 
presented in reference \cite{Rodriguez:2010tn}.

Now, if we use the VEV presented by Eq.(\ref{hip1}) in 
Eqs.(\ref{rhoexpmsusy331},\ref{resmsusy331}), we obtain the 
following results
\begin{eqnarray}
\rho &=& 1 \nonumber \\
\delta M^{2}_{Z}&=&\frac{\delta M^{2}_{W}}{\cos^{2}\theta_{W}} \Rightarrow 
\delta M^{2}_{Z}=17.187592 \,\ {\rm GeV}^{2}, 
\label{zresultfromhip1}
\end{eqnarray}
if we impose the new measurement of the mass of 
the $W$-boson, presented by the CDF collaboration, this 
implies that the experimental measurement for the mass of the 
neutral $Z$-boson must change to
\begin{eqnarray}
\left( M_{Z} \right)_{\rm CDF}&=&91.19 \,\ {\rm GeV.}
\label{mzhip1values} 
\end{eqnarray} 
which are exactly the same results that we had already obtained for the M331, 
review our Eqs.(\ref{dmzdmwm331},\ref{mzcdfvaluesm331}). Therefore, using
\begin{eqnarray}
v_{\sigma_{1}}&=&v_{\sigma^{\prime}_{1}}=0 \,\ 
{\rm GeV,} \nonumber \\
v_{\sigma^{\prime}_{2}}&\simeq&11.19 \,\ {\rm GeV.}
\label{mzhip1}
\end{eqnarray}
is a possible solution to simultaneously explain the value 
$(M_{W})_{\rm CDF}$ and $\rho=1$ in our model and we can predict what 
change should occur in the mass of the $Z$-boson, according to 
Eq.(\ref{mzhip1values}). However, this is not the only solution to 
those problems in this model.

The second hypothesis we want to present, we will define the 
VEV coming from the sextet are zero, it means
\begin{equation}
v_{\sigma^{\prime}_{1}}=v_{\sigma^{\prime}_{2}}=0 \,\ 
{\rm GeV},
\label{hip2}
\end{equation}
in this case we recovered the M331. However, in this case, the
Eqs.(\ref{limitrhom331},\ref{limitMWm331}) remain valid and 
therefore in this case there is no solution for both problems 
under this hypothesis within MSUSY331.
Therefore we will discard it as we did in the context of M331, please see our 
introduction, and to solve the problem in the mass of the
$W$-boson was given by reference \cite{VanLoi:2022eir}.   

For the third hypothesis, we will choose that all new VEV of 
this model are different from zero, this means
\begin{equation}
v_{\sigma_{1}}\neq 0, v_{\sigma^{\prime}_{1}}\neq 0, 
v_{\sigma^{\prime}_{2}}\neq 0.
\end{equation}
Our main objective in this article is to simultaneously 
explain the experimental value of the CDF for $M_{W}$ and 
also $\rho=1$.

\begin{figure}[ht]
\begin{center}
\vglue -0.009cm
\mbox{\epsfig{file=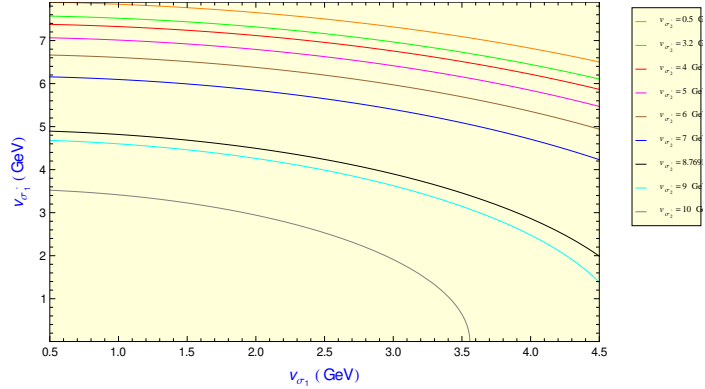,width=0.7\textwidth,angle=0}}       
\end{center}
\caption{Possible values for $v_{\sigma^{\prime}_{1}}$ 
as function of $v_{\sigma_{1}}$, in the minimal 
supersymmetric  $SU(3)_{C}\otimes SU(3)_{L}\otimes U(1)_{N}$ 
Model,  see Eq.(\ref{sigma1VSsigma1p}), for some values of 
$v_{\sigma^{\prime}_{2}}$, the reason why we choose some of those value 
see the comments below Eq.(\ref{expvalmassdiff}).}
\label{fig6}
\end{figure}

We will begin our numerical analysis by rewriting Eq.(\ref{resmsusy331}) as follows
\begin{eqnarray}
v^{2}_{\sigma^{\prime}_{2}} +2 \left(
v^{2}_{\sigma^{1}}+v^{2}_{\sigma^{\prime}_{1}} \right)= 
\frac{4M^{2}_{W}}{g^{2}}, 
\label{eqfoda}
\end{eqnarray} 
however, Eq.(\ref{rhoexpmsusy331}), allows us to write the 
following equality
\begin{eqnarray}
2 \left( \frac{
v^{2}_{\sigma^{1}}+v^{2}_{\sigma^{\prime}_{1}}}{v^{2}_{\rm SM}} 
\right)= 
0.0008.
\end{eqnarray}
We can use both equations for fix $v_{\sigma^{\prime}_{2}}$ 
in the following way
\begin{eqnarray}
v_{\sigma^{\prime}_{2}}&=& 
\sqrt{\frac{4\delta M^{2}_{W}}{g}-0.0008*v^{2}_{\rm SM}}=
8.7691841 \,\ {\rm GeV}.
\label{vsigma2phip3}
\end{eqnarray}
Using Eq.(\ref{eqfoda}) we can define $v_{\sigma_{1}}$ to be 
expressed through the following algebraic expression
\begin{equation}
v_{\sigma^{\prime}_{1}}= \sqrt{\frac{2\delta M^{2}_{W}}{g}- 
\frac{v^{2}_{\sigma^{\prime}_{2}}}{2}-v^{2}_{\sigma_{1}}}
\label{sigma1VSsigma1p}
\end{equation}
our results is shown in Fig.(\ref{fig6}), where we see for 
small values of $v_{\sigma^{\prime}_{2}}<7$ GeV, we get 
$v_{\sigma^{\prime}_{1}}>5$ GeV. 

However, we want both VEVs, $v_{\sigma_{1}},v_{\sigma^{\prime}_{1}}$, to have values ​​that 
are close to or less than $5$ GeV, for this we will use numerical values presented by Fig.(\ref{fig6}). 
As a conclusion of this numerical analyses, given in our 
Tab.(\ref{tabzsector}), we can say
\begin{eqnarray}
\rho &\simeq&0.9992 \nonumber \\ 
\delta M^{2}_{Z}&\simeq&23.83 \,\ {\rm GeV}^{2}, \,\ \Rightarrow 
\left( M_{Z} \right)_{\rm CDF}\simeq 91.32 \,\ {\rm GeV} .
\label{mzhip3}
\end{eqnarray} 

\begin{table}
\begin{tabular}{|c|c|c|c|c|}
\hline
$v_{\sigma_{1}}$ GeV & $v_{\sigma^{\prime}_{1}}$ GeV & 
$\rho$ & $\delta M^{2}_{Z}$ GeV 
& $(M_{Z})_{CDF}$ GeV  \\ 
\hline
0.5 & 4.89453 & 0.992 & 23.8278 & 91.3181 \\
1.0 & 4.81730 & 0.992 & 23.8278 & 91.3181 \\
1.5 & 4.68577 & 0.992 & 23.8278 & 91.3181 \\
2.0 & 4.49515 & 0.992 & 23.8278 & 91.3181 \\
2.5 & 4.2375 & 0.992 & 23.8278 & 91.3181 \\
3.0 & 3.89954 & 0.992 & 23.8278 & 91.3181 \\
3.5 & 3.4578 & 0.992 & 23.8278 & 91.3181 \\
4.0 & 2.86468 & 0.992 & 23.8278 & 91.3181 \\
4.5 & 1.98907 & 0.992 & 23.8278 & 91.3181 \\
\hline
\end{tabular}
\caption{Defined some values of $v_{\sigma_{1}}$ we show the numerical values for $v_{\sigma^{\prime}_{1}}$ GeV; 
$\rho$-parameter and the values of 
$(M_{Z})_{CDF}$ in agreement with our 
Fig.(\ref{fig6}) and using the VEV of 
$\sigma^{\prime}_{2}$ given in Eq.(\ref{vsigma2phip3}).}
\label{tabzsector}
\end{table}

We will consider for charged gauge bosons only the values ​​of
$v_{\sigma_{1}}$ and $v_{\sigma^{\prime}_{1}}$ given by
first line of our Tab.(\ref{tabzsector}), for the other values
presented in this the masses can be obtained in a similar way. Now 
the gauge bosons $W_{1}$ and $W_{2}$ are the physical ones. Our results can 
be found in Tabs.(\ref{tab2b},\ref{tab3},\ref{tab4}) and the general 
concluison are
\begin{itemize}
\item $M_{W_{1}}\cong M_{W}$ and $M_{W_{2}}\cong M_{V}$
\item $M_{Z_{1}}\cong M_{Z}$ and $M_{Z_{2}}\cong M_{Z^{\prime}}$
\item $M_{Z_{2}}>M_{U}>M_{W_{2}}$.
\end{itemize}
Note that for small values ​​of $v_{\chi}$ we ​​have 
$M_{U}>M_{V}$, see Tab.(\ref{tab2b}), whereas for large values ​​of $v_{\chi}$ 
the two charged gauge bosons are almost degenerate, it means 
$M_{U}\simeq M_{V}$ as shown in Tab.(\ref{tab4}) and see also our 
Eq.(\ref{bgmcc}) to really convince themselves of these facts. 

\begin{table}
\begin{tabular}{|c|c|c|c|c|c|c|}
\hline
$v_{\chi^{\prime}}$ GeV & $M_{W}$ GeV & $M_{V}$ GeV 
& $\delta_{WV}$ GeV$^{2}$ & $M_{W_{1}}$ GeV & $M_{W_{2}}$ 
GeV & $M_{U}$ GeV  \\ 
\hline
1000 & 80.4325 & 461.582 & 14.2305 & 80.4335 & 461.582 
& 468.481 \\
1500 & 80.4325 & 588.368  & 14.2305 & 80.4335 & 588.368 
& 593.797 \\
2000 & 80.4325 & 729.757  & 14.2305 & 80.4335 & 729.757 
& 734.140 \\
2500 & 80.4325 & 878.726  & 14.2305 & 80.4335 & 878.726 
& 882.370 \\
3000 & 80.4325 & 1032.00  & 14.2305 & 80.4335 & 1032.00 
& 1035.10 \\
3500 & 80.4325 & 1187.91  & 14.2305 & 80.4335 & 1187.91 
& 1190.61 \\
4000 & 80.4325 & 1345.55  & 14.2305 & 80.4335 & 1345.55 
& 1347.93 \\
4500 & 80.4325 & 1504.36  & 14.2305 & 80.4335 & 1504.36 
& 1506.49 \\
5000 & 80.4325 & 1664.02  & 14.2305 & 80.4335 & 1664.02 
& 1665.94 \\
\hline
\end{tabular}
\caption{Masses of charged gauge bosons as function of $v_{\chi^{\prime}}$ 
and $v_{\chi}=1000$ GeV, the others VEVs are shown at 
Eqs.(\ref{standardvevm331},\ref{fixrho},\ref{hip1}), see the textet and 
those values are in agreement with Eq.(\ref{explimV}), but these values 
​​do not satisfy the constraint given by Eq.(\ref{explimV1}).}
\label{tab2b}
\end{table}

\begin{table}
\begin{tabular}{|c|c|c|c|c|c|}
\hline
$v_{\chi^{\prime}}$ GeV & $M_{Z}$ GeV & $M_{Z^{\prime}}$ GeV 
& $\delta_{ZZ^{\prime}}$ GeV$^{2}$ & $M_{Z_{1}}$ GeV & $M_{Z_{2}}$ 
GeV  \\ 
\hline
1000 & 91.3181 & 2023.91 & -21956.4 & 91.2817 & 2023.91   \\
1500 & 91.3181 & 2579.43 & -21956.4 & 91.2817 & 2579.43  \\
2000 & 91.3181 & 3199.01 & -21956.4 & 91.2817 & 3199.01  \\
2500 & 91.3181 & 3851.85 & -21956.4 & 91.2817 & 3851.85  \\
3000 & 91.3181 & 4523.57 & -21956.4 & 91.2817 & 4523.57  \\
3500 & 91.3181 & 5206.88 & -21956.4 & 91.2817 & 5206.88  \\
4000 & 91.3181 & 5897.75 & -21956.4 & 91.2817 & 5897.75  \\
4500 & 91.3181 & 6593.79 & -21956.4 & 91.2817 & 6593.79  \\
5000 & 91.3181 & 7293.54 & -21956.4 & 91.2817 & 7293.54  \\
\hline
\end{tabular}
\caption{Masses of physical neutral gauge bosons as function of 
$v_{\chi^{\prime}}$ and $v_{\chi}=1000$ GeV, the others VEVs are shown at 
Eqs.(\ref{standardvevm331},\ref{fixrho},\ref{hip1}), see the textet and 
those values are in agreement with Eq.(\ref{explimZP}) and more detail in 
App.(\ref{zezp}).}
\label{tab3}
\end{table}

\begin{table}
\begin{tabular}{|c|c|c|c|c|c|}
\hline
$v_{\chi^{\prime}}$ GeV & $M_{V}$ GeV & $M_{W_{2}}$ GeV & 
$M_{U}$ Gev & $M_{Z^{\prime}}$ TeV & $M_{Z_{2}}$ TeV  \\ 
\hline
12.0 & 5538.12 & 5538.12 & 5538.70 & 24.27 & 24.27 \\
12.5 & 5654.68 & 5654.68 & 5655.24 & 24.78 & 24.78 \\
13.0 & 5773.49 & 5773.49 & 5774.05 & 25.31 & 25.31 \\
13.5 & 5894.43 & 5894.43 & 5894.97 & 25.84 & 25.84  \\
14.0 & 6017.36 & 6017.36 & 6017.89 & 26.37 & 26.37   \\
14.5 & 6142.16 & 6142.16 & 6142.68 & 26.92 & 26.92   \\
15.0 & 6268.73 & 6268.73 & 6269.24 & 27.48 & 27.48   \\
15.5 & 6396.95 & 6396.95 & 6397.45 & 28.04 & 28.04   \\
\hline
\end{tabular}
\caption{Masses of heavy gauge bosons as function of 
$v_{\chi^{\prime}}$ and $v_{\chi}=12$ TeV, the others VEVs are shown at 
Eqs.(\ref{standardvevm331},\ref{fixrho},\ref{hip1}), those values are 
in agreement with Eqs.(\ref{explimV1},\ref{explimZP}). The masses of $W$ and $W_{1}$ are the same as shown in Tab.(\ref{tab2b}), for the case of $Z$ and  $Z_{1}$ their values are provided in our Tab.(\ref{tab3}).}
\label{tab4}
\end{table}

We thought it would be useful to analyze the phenomenological studies 
presented by references 
\cite{dutta,Pankov:2019yzr,Osland:2020onj,barela,barela1} in 
this supersymmetric model and in this way see if the bonds are maintained 
or not. Also analyze at which scale this model loses its perturbative 
characteristic in accordance with the Eq.(\ref{barelalimite}) presented by 
the reference \cite{Barela:2023oyp}. It is of fundamental importance to 
see whether, using the analyzes presented here, how the masses of the 
lightest scalar in this model change, that is, whether the 
Eq.(\ref{higgsmassMSUSY331}) presented in the references 
\cite{331susy1,esc2} still holds.

\section{Conclusions}
\label{sec:con}

We made a detailed study on the masses as well as the mixture of all 
gauge bosons in the Minimal Supersymmetric 
$SU(3)_{C}\otimes SU(3)_{L}\otimes U(1)_{N}$ Model, MSUSY331. 
Our first result, we show how allowing all non-zero VEV of both 
the anti-sextet and sextet, we achieve to explain both the new 
measurement on the $W$-boson mass as well as the $\rho$-parameter if 
Eqs.(\ref{wmassrest},\ref{rhorest}) are satisfied 
simultaneously as we have showed in 
Eqs.(\ref{mzhip1},\ref{mzhip3}) and also in our 
Fig.(\ref{fig6}) and Tabs.(\ref{tab2b},\ref{tab3},\ref{tab4}). 
The new values ​​for the $Z$-boson masses to be in agreement 
with the new mass result for $W$-boson mass presented by 
Fermilab's CDF collaboration are presented by 
Eq.(\ref{mzhip1values}) and the second line in our 
Eq.(\ref{mzhip3}). We also show as a general conclusion that the 
masses of all measurement bosons of this model satisfy the following 
hierarchy in their masses $M_{Z^{\prime}}>M_{U}>M_{V}$ and the 
numerical values ​​obtained are in agreement
with the actual experimental data provided in
Eqs.(\ref{explimV},\ref{explimV1},\ref{lowerlimitU},\ref{explimZP}).

\begin{center}
{\bf Acknowledgments} 
\end{center}
We would like thanks V. Pleitez for useful discussions above 331 models and 
about this intersting research topic. We also to thanks IFT for the nice hospitality during 
my several visit to perform my studies about the severals 331 Models and also for done 
this article.

\appendix

\section{Mixing between $W$ and $V$.}
\label{sec:mixingchar}

We notice, from Eq.(\ref{bgmcc}), 
if $v_{\sigma_{1}}\neq 0$ the 
gauge bosons $W$ and $V$ are no longer the physical ones, in 
agreement with the results presented at \cite{VanLoi:2022eir,Liu:1993gy,Liu:1993fwa,Montero:1999mc}. They 
can mixing and the physical bosons are $W^{\pm}_{1,2}$ and their masses are
\begin{eqnarray}
M^{2}_{W_{1}}&=& \frac{1}{2} \left( 
M^{2}_{W}+M^{2}_{V}- \sqrt{(M^{2}_{W}-M^{2}_{V})^{2}+4 \delta^{2}_{WV}} 
\right), \nonumber \\
M^{2}_{W_{2}}&=& \frac{1}{2} \left( 
M^{2}_{W}+M^{2}_{V}+ \sqrt{(M^{2}_{W}-M^{2}_{V})^{2}+4 \delta^{2}_{WV}} 
\right).
\label{eigenvaluescc}
\end{eqnarray}
The physical eigenstates are defined as
\begin{eqnarray} 
\left( \begin{array}{c}
W^{\pm}_{1m}\\
W^{\pm}_{2m} \end{array} \right)= \left( \begin{array}{cc}
\cos \theta^{\pm}&- \sin \theta^{\pm}\\
\sin \theta^{\pm}& \cos \theta^{\pm} 
\end{array} 
\right) \left( 
\begin{array}{c}
W^{\pm}_{m}\\
V^{\pm}_{m} \end{array}\right),
\label{eigenvectorscc}
\end{eqnarray}
the mixing angle is given by
\begin{equation}
\tan^{2} \theta^{\pm}= 
\frac{M^{2}_{ W_{1}}-M^{2}_{W}}{M^{2}_{ W_{2}}-M^{2}_{V}}.
\label{thetac}
\end{equation}

\section{Mixing between $Z$ and $Z^{\prime}$.}
\label{sec:mixingneu}

The neutral gauge bosons $Z$ and $Z^{\prime}$ can mix and it 
is given by
\begin{eqnarray} 
{\cal L}^{{\rm neutra}}&=& \left( \begin{array}{cc}
Z_{m}& Z^{\prime}_{m} \end{array} \right) \left( \begin{array}{cc}
M^{2}_{Z}& \delta_{ZZ^{\prime}} \\
\delta_{ZZ^{\prime}}& M^{2}_{Z^{\prime}} \end{array} \right) \left( \begin{array}{c}
Z_{m}\\
Z^{\prime}_{m} \end{array} \right),
\label{neutralbgmcc}
\end{eqnarray}
where 
\begin{eqnarray}
M^{2}_{Z}&=& \frac{g^{2}v^{2}_{MP}}{4 \cos^{2} \theta_{W}}, \nonumber \\
M^{2}_{Z^{\prime}}&=& \frac{g^{2}}{4}\left[ \frac{4v^{2}_{MP}}{3s}+ \frac{2t^{2}U^{2}}{s}+ \frac{4s}{3} 
\left( w^{2}+ w^{\prime 2} \right) \right], \nonumber \\
\delta_{ZZ^{\prime}}&=& \frac{g^{2}}{4 \sqrt{3h_{w}}s}\left[ (V^{2}-U^{2})-6t^{2}U^{2} \right],
\label{zmass}
\end{eqnarray}
Through comparing those values with our result presented on 
Eq.(\ref{zmassmcr}), we have defined
\begin{eqnarray}
V^{2}&=&v^{2}+2y^{2}+z^{2}+v^{\prime 2}+2y^{\prime 2}+z^{\prime 2}, \nonumber \\
U^{2}&=&u^{2}+ u^{\prime 2}, \nonumber \\
t^{2}&=& \frac{\sin^{2}\theta_{W}}{1-4 \sin^{2}\theta_{W}}, \nonumber \\
s&=&1+3t^{2}, \nonumber \\
v^{2}_{\rm MP}&=&V^{2}+U^{2},
\end{eqnarray}
To understand how to get the third expression above see, Eqs.(\ref{tinthetaw},\ref{tinthetaw2}). 
We can show the following relations
\begin{eqnarray}
\cos \theta_{W}&=& \frac{\sqrt{s}}{\sqrt{s+t^{2}}}, \nonumber \\
\sin \theta_{W}&=& \frac{t}{\sqrt{s+t^{2}}}, \nonumber \\
\sqrt{s+t^{2}}&=&\left( \sqrt{h_{W}} \right)^{-1}, \nonumber \\
h_{W}&=&1-4 \sin^{2}\theta_{W}.
\label{def:hw}
\end{eqnarray}

The physical mass of physical gauge bosons are
\begin{eqnarray} 
\left( \begin{array}{c}
Z^{0}_{1m}\\
Z^{0}_{2m} \end{array} \right)= \left( \begin{array}{cc}
\cos \theta^{0}&- \sin \theta^{0}\\
\sin \theta^{0}& \cos \theta^{0} 
\end{array} 
\right) \left( 
\begin{array}{c}
Z^{0}_{m}\\
(Z^{\prime})^{0}_{m} \end{array}\right),
\label{eigenvectorscc}
\end{eqnarray}
the mixing angle in this sector is
\begin{equation}
\tan^{2} \theta^{0}= 
\frac{M^{2}_{ Z_{1}}-M^{2}_{Z}}{M^{2}_{ Z_{2}}-M^{2}_{Z^{\prime}}},
\label{thetac}
\end{equation}
where
\begin{eqnarray}
M^{2}_{Z_{1}}&=& \frac{1}{2} \left( 
M^{2}_{Z}+M^{2}_{Z^{\prime}}+ \sqrt{(M^{2}_{Z}-M^{2}_{Z^{\prime}})^{2}+4 
\delta^{2}_{ZZ^{\prime}}} 
\right), \nonumber \\
M^{2}_{Z_{2}}&=& \frac{1}{2} \left( 
M^{2}_{Z}+M^{2}_{Z^{\prime}}- \sqrt{(M^{2}_{Z}-M^{2}_{Z^{\prime}})^{2}+4 
\delta^{2}_{ZZ^{\prime}}} 
\right).
\label{eigenvaluesnc}
\end{eqnarray}

\section{Preliminar Numerical Analysis for $Z$ and $Z^{\prime}$}
\label{zezp}

We are going to use the VEV defined at 
Eqs.(\ref{standardvevm331},\ref{fixrho}), and as an example as the 
mixing of $Z$ and $Z^{\prime}$, together the following VEV
\begin{eqnarray}
v_{\sigma_{1}}&=&v_{\sigma^{\prime}_{1}}=
v_{\sigma^{\prime}_{2}}=0 \,\ {\rm GeV}, \nonumber \\
v_{\chi}&=&v_{\chi^{\prime}}=1000 \,\ {\rm GeV},
\end{eqnarray}
Eq.(\ref{zmass}) can get the following results\footnote{The 
parameter $\delta_{ZZ^{\prime}}$ has no dependence in 
$v_{\chi}$ and $v_{\chi^{\prime}}$, see Eq.(\ref{zmass}) and also our 
Tab.(\ref{tab3}).}
\begin{eqnarray}
M_{Z}&=&91.1875 \,\ {\rm GeV}, \,\
M_{Z^{\prime}}=2023.91 \,\ {\rm GeV} \nonumber \\
\delta_{ZZ^{\prime}}&=&-21958 \,\ {\rm GeV}^{2},
\label{ZSM}
\end{eqnarray}
the value for $Z^{\prime}$ mass is in agreement with Eq.(\ref{explimZP}). Repare $\delta_{ZZ^{\prime}} \neq 0$, therefore $Z$ and 
$Z^{\prime}$ are not physical gauge boson. The masses of 
physical gauge bosons $Z_{1,2}$, using 
Eq.(\ref{eigenvaluesnc}), are\footnote{Compare 
$\delta_{ZZ^{\prime}}$ with $M^{2}_{Z}$ and 
$M^{2}_{Z^{\prime}}$ and it is smaller therefore the mixing 
parameter is almost negligible.}
\begin{eqnarray}
M_{Z_{1}}&=&91.1875 \,\ {\rm GeV}, \,\
M_{Z_{2}}=2023.91 \,\ {\rm GeV}.
\end{eqnarray}
Therefore the bosons $Z$ and $Z^{\prime}$, as a first approximation, can be considered 
as being the physical states, and their masses given at 
Eq.(\ref{zmass}). We get the same conclusion in our Tab.(\ref{tab3}).


\end{document}